\documentclass[aps,pre,twocolumn,superscriptaddress]{revtex4-1}
\usepackage[T1,T2A]{fontenc}
\usepackage[utf8]{inputenc}
\usepackage[russian,english]{babel}
\usepackage{graphicx}
\usepackage{dcolumn}
\usepackage{bm}
\usepackage{graphicx}
\usepackage{amsmath}
\usepackage{amssymb}
\usepackage{pifont}
\usepackage{epstopdf}
\usepackage{color}

\newcommand{\Da}{\textrm{Da}}

\begin{document}

\title{Electrophoretic velocity
of ion-releasing colloidal particles}

\author{Evgeny S. Asmolov}
\affiliation{Frumkin Institute of Physical Chemistry and Electrochemistry, Russian Academy of Sciences, 31 Leninsky Prospect, 119071 Moscow, Russia}
\author{Victoria A. Vasileva }
\affiliation{Frumkin Institute of Physical Chemistry and Electrochemistry, Russian Academy of Sciences, 31 Leninsky Prospect, 119071 Moscow, Russia}
\author{Olga I. Vinogradova}
\email[Corresponding author: ]{oivinograd@yahoo.com}
\affiliation{Frumkin Institute of Physical Chemistry and Electrochemistry, Russian Academy of Sciences, 31 Leninsky Prospect, 119071 Moscow, Russia}

\begin{abstract}
By means of a matched asymptotic expansions approach the electrophoretic velocity and zeta potential of a catalytic particle that uniformly releases ions have been investigated. Attention is focused on large, compared to diffuse layer, particles characterized, beside the surface potential $\Phi_s$, by the Damk\"{o}hler number Da that represents the
ratio of the surface reaction rate to the diffusive transfer one. For vanishing Da, we recover the classical Smoluchowski formula for the electrophoretic velocity which states that the zeta potential of the particle is equal to $\Phi_s$ and that the migration direction is determined by its sign. For small values of Da we show that the migration velocity is controlled mostly by $\Phi_s$ and affected by an ion release only slightly. However, even small Da can induce the electrophoresis of electro-neutral particles that would be immobile if inert. For larger Da the direction of migration and the sign of zeta potential become independent on  $\Phi_s$ and are solely determined by the difference in diffusivity of released cations and anions. Still,  the surface potential affects the magnitude of the particle velocity.
\end{abstract}

	\date{\today}
	\maketitle

\section{Introduction}

When an electric field $E$ is applied to a colloidal solution, the particles
of radius $R$ migrate with a constant speed $U$ to one of the electrodes.
This phenomenon, termed electrophoresis, takes its origin in the
electrostatic diffuse layer (EDL) formed close to the charged particle,
which extends to distances comparable to the Debye length $\lambda_D$ of a
bulk solution. Electrophoresis represents a complex phenomenon controlled by (low Reynolds
number) hydrodynamics, electrostatics, and ionic transport. Its physical origin is more or less understood. A tangential electric field generates inside the EDL a force
that sets the fluid in motion. The emergence of such a flow in turn provides
hydrodynamic stresses that cause the steady-state propulsion of the particle.

Electrophoretic experiments have played a great role in the development of
colloid science~\cite{overbeek}. This is due to the fact that there is a
number of techniques for the determination of the electrophoretic velocity
that require a relatively simple experimental devices. Another factor in
favor of electrophoresis certainly has been its simple interpretation by Smoluchowski~\cite%
{smoluchowski.m:1921}, permitting
to infer the so-called electrokinetic or zeta potential $Z$ from the measured electrophoretic mobility:
\begin{equation}
\dfrac{U}{E} =\dfrac{\epsilon }{4\pi \eta }Z,
\label{eq:smoluchowsky}
\end{equation}%
where $\epsilon$ is the dielectric constant and $\eta$ is the dynamic viscosity. Smoluchowski argued that $Z$ is completely independent on the particle radius, if large enough: $\lambda =\lambda _{D}/R\ll 1$, and must be equal to the surface (electrostatic) potential $\Phi_s$ provided that the (hydrodynamic) no-slip is postulated. Indeed simple electrokinetic experiments are often used to
invoke the surface potential and interpret phenomena it determines. For
instance, the stability of colloidal suspensions was found to be closely
related to $Z$ calculated from electrophoretic data, and such data are much
easier to obtain compared to direct surface force measurements~\cite{israelachvili.jn:2011}.

Nowadays, beside colloid science, the electrophoresis finds numerous applications in
many areas, such as microfluidics, DNA sequencing, drug delivery, analytical
chemistry, and medicine~\cite{squires.tm:2005,ruiz.mc:1993,chen.st:2020}, and the field continues to advance rapidly. Throughout the years,
there have been many attempts to provide a more satisfactory theory of colloid electrophoresis. Note that, if $\lambda _{D}$ is much smaller than the interparticle separation, the space between the colloids is electro-neutral, and this region is force-free (no flow). By this reason, the problem of electrophoresis of
dilute colloidal suspensions can normally be treated on a single-particle level. We mention below what we believe are the more relevant contributions, concentrating on the case of a single nonconducting sphere.

It is now well recognized that if we make the assumption of a weak applied field, the electrophoresis is linear ($U \propto E$) and \eqref{eq:smoluchowsky} represents the general equation for an electrophoretic mobility, valid for any particle. A main task is thus to calculate the zeta potential, which obviously depends on the radius of curvature and hydrodynamic boundary conditions  at the particle surface.
Early work has concentrated on the extension the Smoluchowski theory for an arbitrary $\lambda$, and some useful approximate solutions have been proposed~\cite{huckel.e:1924,henry.dc:1931}. As recently turned out, they can even be employed for a description of  electrophoresis of small inorganic ions that determines the conductivity of electrolyte solutions~\cite{vinogradova.oi:2023b}. However, the quantitative understanding of electrophoresis in the general case of
arbitrary $\lambda$ and  $\Phi_s$  is still challenging  due to non-linearity of electrostatic equations and this problem has been solved only numerically~\cite{obrien:1978}. The more recent attempts at extension on the Smoluchowski theory have included the effects of a hydrodynamic slip~\cite{khair.as:2009,ohshima.h:2019} and of a mobility of
adsorbed charges~\cite{vinogradova.oi:2023}. During the last  decades some theoretical papers have been concerned with the nonlinear
electrophoresis in a strong field, and numerical approaches have mostly been followed~\cite{dunweg.b:2008,giupponi.g:2011,khair.as:2022}.

While significant progress in understanding electrophoresis has been made,  existing work  has been focused only on the classical case of inert particles.  The particles, however, can be catalytic, which here means that they  release ions from their surface. The most known example of such colloids is so-called catalytic swimmers that self-propel in electrolyte solutions thanks to an inhomogeneous ion release from their surface~\cite{asmolov2022COCIS}.
Such ionic catalytic swimmers represent active colloids since the origin of their propulsion is related to self-electrophoresis or self-diffusiophoresis~\cite{moran2017phoretic,peng2022generic}. However, a variety of colloid particles can generate an uniform ionic flux from their
surface. The examples of such particles include calcium carbonate micropumps
dissolving in water~\cite{mcdermott2012}, porous microparticles in
illuminated with an appropriate wavelength solutions of a photosensitive
ionic surfactant~\cite{feldmann.d:2020}, or colloids generating a catalytic
reaction at the uniform surface~\cite%
{dey2015micromotors,ma2016enzyme,patino2018fundamental}. Being uniform, such particles are passive and cannot generate the self-propulsion. However, the ability to release ions can dramatically change their migration in even weak external fields. For instance, it has been
recently shown that an uniform ion release affects the
diffusiophoresis of colloid particles~\cite{asmolov.es:2024}, but we are
unaware of any previous work that has addressed the question of the
electrophoretic migration of such uniformly catalytic particles.

In this paper we develop a theory that describes electrophoretic migration of homogeneous catalytic particles that uniformly release ions. Our theory is applicable in the limit of thin EDL and for a weak external electric field, but we make no additional assumptions about other parameters.

Our paper is arranged as follows: In Sec.~\ref{sec:general} some general considerations concerning electrophoresis of passive catalytic particles are presented. Section~\ref{sec:potential} describes the theory of an electrostatic potential around a catalytic particle. Section~\ref{sec:velocity} contains calculations of the velocity of electrophoretic migration. We conclude in Sec.~\ref{sec:conclusion} with a discussion of our results. Appendix~\ref{app} contains the derivation of the equation that determines the EDL contribution to the particle electrophoretic velocity.

\section{General considerations}
\label{sec:general}

\begin{figure}[h]
\begin{minipage}[h]{0.7\linewidth}
\includegraphics[width=0.9\columnwidth ]{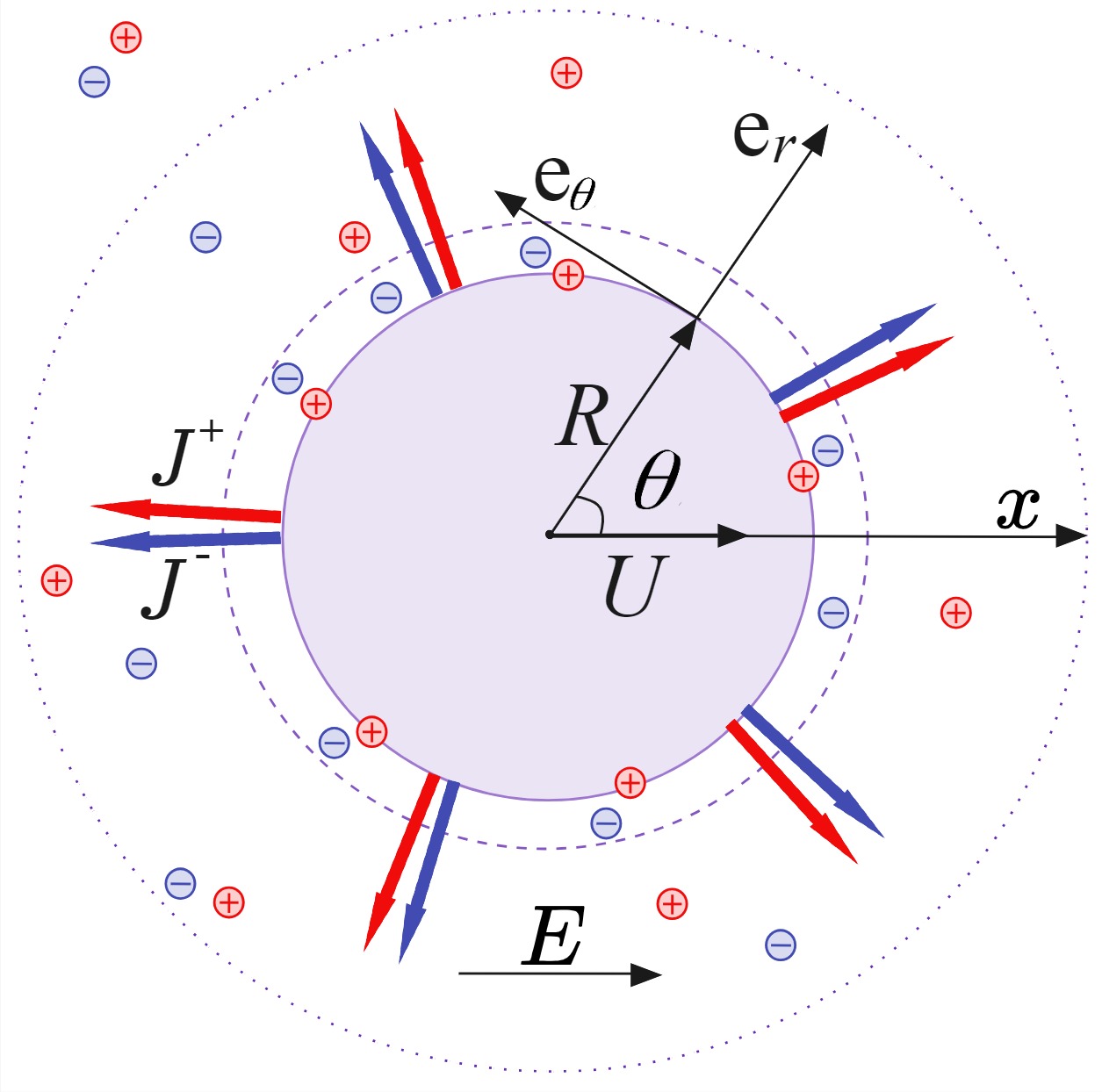} 
\end{minipage}
\vfill
\caption{Schematic representation of a positively charged catalytic particle of radius $R$ migrating with a speed $U$ under an applied electric field $E$. The particle uniformly release ions, and the surface fluxes of anions $J^-$ and cations $J^+$ are equal.
The boundary between the inner region (which is the EDL of thickness $\lambda _{D}$) and the outer one is shown by the dashed curve. The dotted curve illustrates the extension of the weakly charged outer region. }
\label{fig:sketch}
\end{figure}

We consider a spherical particle of radius $R$ immersed in a symmetric $%
\mathrm{z:z}$ aqueous electrolyte solution and exposed to an externally
applied electric field $E$ in $x-$direction (see Fig.~\ref{fig:sketch}). The
Debye screening length of a solution, $\lambda _{D}=\left( 8\pi \mathrm{z}%
^2 \ell _{B} C_{\infty }\right) ^{-1/2}$, where $\ell _{B}$ is the Bjerrum length and $C_{\infty }$ is a number
density (concentration) of ions in the bulk. The EDL is considered to be
thin compared to radius of curvature of the surface, $\lambda = \lambda_D/R \ll 1$,
and the particle is non-conducting, so only a tangential to the surface
component of $E$ induces the migration.

In the absence of the applied field $E$ close to the particle a cloud of diffuse ions of a local volume charge density $\varrho$ is formed. Correspondingly, the dimensionless electrostatic potential $\varphi$ (scaled by $k_B T/\mathrm{e}$, where $k_{B}$ is the Boltzmann
constant, $T$ is a temperature, and $\mathrm{e}$ is the proton charge) is a function of the radial coordinate only. We require the potential and the electric field to
vanish at infinity, while on the spherical surface $\varphi =\phi_{s}$, where $\phi_{s} = \mathrm{e} \Phi_s/(k_B T)$. For inert particles the radial potential profile can be found numerically from the non-linear Poisson-Boltzmann equation. If the latter can be linearized, the local potential is given by the Debye-H\"{u}ckel approximation
\begin{equation}  \label{eq:DH}
\varphi = \phi_s \dfrac{e^{-(r-1)/\lambda}}{r},
\end{equation}
where the radial coordinate $r$ is scaled by $R$.

The classical assumption of an inert particle is relaxed here and we treat the
latter as catalytic. This implies that it can uniformly emit ions. The
released cations and anions are assumed to be the same as in the bulk, and
their fluxes are equal. Clearly, this should affect $\varrho$ and $\varphi$, but not  $\phi_{s}$. Note that the uniform catalytic particle remains
passive, i.e. cannot generate a self-propulsion. The migration, if any,
occurs only under an applied weak field.

The weak electric field  applied in the $x-$direction to produce electrophoresis is additively superimposed upon
the radial field generated by the particle itself. That the electrophoresis of catalytic particles should differ from that of
inert one becomes self-evident from the arguments presented below.

Since the surface
fluxes of anions and cations are equal, the total particle charge remains
constant. The released ions diffuse to the bulk, but their diffusion
coefficients $D^{+}$ and $D^{-}$ are generally unequal, and their difference
can be characterized by the parameter \cite{prieve1984motion,velegol2016}
\begin{equation}
\beta =\frac{D^{+}-D^{-}}{D^{+}+D^{-}}.  \label{bet}
\end{equation}%
In such a definition $\beta$ is positive, if cations diffuse faster than
anions, and vice versa. The values of $\beta$ are typically confined between
$-1$ to $1$. For instance, in the case of salt grain dissolution, $\beta
\simeq 0.285$ for KCH$_3$COO, $0.014$ for KNO$_3$, $-0.207$ for NaCl, and $%
-0.506$ for CaCO$_3$~\cite{velegol2016}. The products of the enzymatic
decomposition of urea at the surface of functionalized particles form NH$%
_{4}^{+}$ and OH$^{-}$ ions~\cite{de2020self} which yields $\beta \simeq
-0.459.$

A local ionic concentration can approach the steady-state only if the fluxes
are equal not just at the surface, but also at any $r$. To provide this, an
additional spherically symmetric electric field should arise around the
particle that accelerate slower ions and retard faster ones. Clearly, this
field is much weaker than inside the EDL, but the extension of the
non-electroneutral region becomes much larger. Consequently, its impact on
the migration velocity can be significant. Thus, one can expect that the
catalytic particle would migrate faster or slower compared to inert,
depending on the sign of $\beta$. We stress that if $\beta = 0$, an additional radial field is not induced, but the local potential $\varphi$ around  catalytic particles becomes different from that near inert ones.

It is convenient to introduce the Damk\"{o}hler number that represents the
ratio of the surface reaction rate to the diffusive transfer rate~\cite%
{cordova2008,moran2017}:
\begin{equation}
\mathrm{Da}=\frac{j R}{D C_{\infty }},\quad D=\frac{2D^{+}D^{-}}{D^{+}+D^{-}}%
.  \label{Da_def}
\end{equation}%
Here $j$ is a dimensional flux of released ions, and $D$ is a harmonic mean
of the diffusion coefficients. Such a definition of $\mathrm{Da}$ implies
that it is small at high salt and/or when the surface flux is small. By
contrast, low $C_{\infty }$, especially along with a large surface flux,
would provide a large value of $\mathrm{Da}$.

Our central goal is to determine the electrophoretic velocity and zeta
potential of such a catalytic particle. More precisely, we try to answer a
question about the role of $\beta$ and \textrm{Da}, if any, in establishing
the speed of an electrophoretic migration.

The Peclet and the Reynolds numbers are assumed to be small, so that the
convective fluxes of ions and fluid are ignored. The ion fluxes are governed
by dimensionless Nernst-Planck equations,
\begin{equation}
\nabla \cdot \mathbf{j}^{\pm }=0,  \label{NPH}
\end{equation}%
\begin{equation}
\mathbf{j}^{\pm }=-\nabla c^{\pm }\mp c^{\pm }\nabla \varphi .  \label{NP1}
\end{equation}%
Here the upper (lower) sign corresponds to the positive (negative) ions, all
coordinates are scaled by $R,$ and the dimensionless concentrations $c^{\pm }$
are scaled by $C_{\infty }$.

The Poisson equation for the potential reads%
\begin{equation}
\Delta \varphi =-\lambda ^{-2}\frac{c^{+}-c^{-}}{2},  \label{pois}
\end{equation}
where $\lambda \ll 1.$

The fluid flow satisfies the Stokes equations,
\begin{equation}
\mathbf{\nabla \cdot v}=0\mathbf{,\quad }\Delta \mathbf{v}-\mathbf{\nabla }p=%
\mathbf{f}.  \label{NS}
\end{equation}%
Here $\mathbf{v}$ and $p$ are the dimensionless fluid velocity and pressure
(scaled by $\dfrac{k_{B}T}{4\pi \eta \ell _{B}R}$ and $\dfrac{k_{B}T}{4\pi
\ell _{B}R^{2}}$), and
\begin{equation}
\mathbf{f}=-\Delta \varphi \mathbf{\nabla }\varphi ,  \label{v_rec}
\end{equation}%
is the electrostatic body force.

\section{Electrostatic potential}\label{sec:potential}

Since we consider the case of $\lambda \ll 1, $ one can construct the
asymptotic solution involving separate expansions covering the outer and
inner regions with different lengthscales. The inner region is defined
inside the EDL formed close to the particle and its lengthscale is obviously
$\lambda_D$. We identify the outer region with the weakly charged cloud
formed due to a diffusion of released ions that extends to much larger
distances. The inner and outer expansions are then matched asymptotically at
the boundary between the two regions where both expansions are valid and,
together, constitute a perturbation solution that is valid in the entire
domain.

\subsection{Outer solution}

The boundary of the outer region is located at $r=1+\lambda$, but since $%
\lambda \ll 1$ constructing the solution in the large outer region of
thickness $O(1)$ we impose boundary conditions at the particle surface ($r=1$%
).

The boundary conditions for concentrations at $r=1$ set the ion fluxes. In
the spherical coordinate system they can be formulated as:%
\begin{equation}
\partial _{r}c_{o}^{\pm }\pm c_{o}^{\pm }\partial _{r}\varphi _{o}=-\mathrm{%
Da}\frac{D}{D^{\pm }},  \label{bc_1}
\end{equation}%

Far away from the surface ($r\rightarrow \infty $) the concentration takes
its bulk value
\begin{equation}
 c_{o}^{+}=c_{o}^{-}=1,  \label{bci}
\end{equation}%
and a weak $\left( \varepsilon \ll 1\right) $ constant electric field is
acting in the $x-$direction.
\begin{equation}
\mathbf{\nabla }\varphi _{o}=-\varepsilon \mathbf{e}_{x}=-\dfrac{\mathrm{e} E R \mathbf{e}%
_{x}}{k_{B}T}.  \label{bcif}
\end{equation}

Since $\lambda $ is small, the leading-order solution of (\ref{pois}) is $%
c_{o}^{+}=c_{o}^{-}=c_{o},$ i.e. the electroneutrality holds to $O\left(
\lambda ^{2}\right) .$ However, one should take into account the small
charge $c_{o}^{+}-c_{o}^{-}=O\left( \lambda ^{2}\right) $, since it induces
a finite potential difference in the outer region. Nernst-Planck equations
then become
\begin{eqnarray}
\Delta c_{o}+\nabla \cdot \left( c_{o}\nabla \varphi _{o}\right)  &=&0,
\label{NP2} \\
\Delta c_{o}-\nabla \cdot \left( c_{o}\nabla \varphi _{o}\right)  &=&0.
\label{NP3}
\end{eqnarray}%
Summing up and subtracting Eqs.~(\ref{NP2}) and (\ref{NP3}) along with the
boundary conditions (\ref{bc_1}) we obtain \cite%
{nourhani2015,asmolov2022COCIS}:
\begin{eqnarray}
\Delta c_{o} &=&0,  \label{dif} \\
\nabla \cdot \left( c_{o}\nabla \varphi _{o}\right)  &=&0,  \label{pot2}
\end{eqnarray}%
with the boundary conditions at $r=1$:
\begin{gather}
\partial _{r}c_{o}=-\mathrm{Da},  \label{bc1} \\
c_{o}\partial _{r}\varphi _{o}=\beta \mathrm{Da},  \label{bc_p}
\end{gather}

We have to set the value of the surface potential $\phi _{s}$ and
require $\varphi =\phi _{s}$ at $r=1$. However, this condition is inconsistent with (\ref{bc_p}),
and could, obviously, only be satisfied for the inner solution, but not for the outer.

Since the boundary conditions (\ref{bci}) and (\ref{bc1}) to the Laplace
equation (\ref{dif}) are homogeneous its solution can be readily obtained%
\begin{equation}
c_{o}=1+\frac{\mathrm{Da}}{r}  \label{c_sol}
\end{equation}%
We remark and stress that, if $\Da \neq 0$, the concentration field $c_{o}$ is not equal to the bulk value $c_o=1$ since released ions diffuse to the outer region too. Thus Eq.~(\ref{pot2}) for the potential $\varphi_o$ no longer represents the Laplace equation as it would be for inert particles.   As a result, even for $\beta =0$ the potential will be different from that in the classical case of inert particles.

The boundary conditions (\ref{bcif}) and (\ref{bc_p}) to  linear Eq.~(\ref%
{pot2}) imply that the potential can be sought in the form%
\begin{equation}
\varphi _{o}=\varphi _{o}^{(0)}\left( r\right) +\varepsilon \varphi
_{o}^{(1)}\left( r\right) \cos \theta ,  \label{expa}
\end{equation}%
where $\varphi _{o}^{(0)}$ is the spherically symmetric potential around the
particle, and $\varphi _{o}^{(1)}$ is the disturbance potential due to an
applied electric field.

Substituting the last expansion into Eq.~(\ref{pot2}) we obtain ordinary
differential equations governing $\varphi _{o}^{(0)}$ and $\varphi _{o}^{(1)}
$:%
\begin{equation}
\partial _{r}\left[ r^{2}\left( 1+\frac{\mathrm{Da}}{r}\right) \partial
_{r}\varphi _{o}^{(0)}\right] =0,  \label{fi0}
\end{equation}%
\begin{equation}
\partial _{rr}\varphi _{o}^{(1)}+\frac{2r+\mathrm{Da}}{\left( r+\mathrm{Da}%
\right) r}\partial _{r}\varphi _{o}^{(1)}-\frac{2}{r^{2}}\varphi
_{o}^{(1)}=0,  \label{du}
\end{equation}%
with the boundary conditions at $r=1$
\begin{equation}
\partial _{r}\varphi _{o}^{(0)}=\beta \frac{\mathrm{Da}}{1+\mathrm{Da}},%
\mathrm{\quad }\partial _{r}\varphi _{o}^{(1)}=0,  \label{bcs0}
\end{equation}%
and $r\rightarrow \infty $
\begin{equation}
\varphi _{o}^{(0)}=0,\quad \varphi _{o}^{(1)}=-r.  \label{bci0}
\end{equation}

Solution to Eq.(\ref{fi0}) satisfying (\ref{bcs0}) and (\ref{bci0}) is given
by
\begin{equation}
\varphi _{o}^{(0)}=-\beta \ln \left( 1+\frac{\mathrm{Da}}{r}\right) .
\label{eq:phi0_out}
\end{equation}%
Thus the potential $\varphi _{o}^{(0)}$ is linear in $\beta $ and weakly
logarithmically depends on Da. Clearly, if $\beta $ or Da vanishes, $\varphi
_{o}^{(0)}=0$, i.e. it cannot be induced if the diffusion coefficients of
anions and cations are equal, or in the case of inert particles.

The (symmetric) electric field in the outer region reads
\begin{equation}
\mathbf{\nabla }\varphi _{o}^{(0)}=\beta \frac{\mathbf{\nabla }c_{o}}{c_{o}}%
=-\beta \frac{\mathrm{Da}\,\mathbf{e}_{r}}{r\left( r+\mathrm{Da}\right) }.
\label{dfh}
\end{equation}%
The absolute value of the field decays like $r^{-2}$ as $r\rightarrow \infty
$, but when $r=O(1)$ and $\mathrm{Da}$ is large it demonstrates slower decay
$\propto r^{-1}$. The local volume charge density $\varrho
=c_{o}^{+}-c_{o}^{-}$, which generates the outer field around the particle,
can be calculated from (\ref{pois}):%
\begin{equation}
\varrho =2\lambda ^{2}\beta \Delta \ln \left( c_{o}\right) =-2\lambda
^{2}\beta \frac{\mathrm{Da}^{2}}{r^{2}\left( r+\mathrm{Da}\right) ^{2}}.
\label{char0}
\end{equation}%
\begin{figure}[th]
\centering
\includegraphics[width=1\columnwidth ]{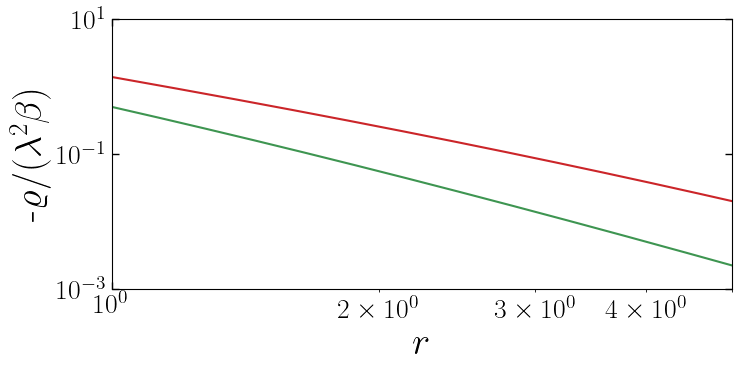} 
\caption{Outer volume charge density calculated from Eq.~\eqref{char0} using
$\text{Da}=5$, $1$ (solid curves from top to bottom). }
\label{fig:char}
\end{figure}

We conclude that $\varrho$ is small compared to the volume charge density in
the inner region being of the order of $\lambda ^{2}\mathrm{Da}^{2}\ll 1,$
but this outer cloud is much larger than inner (i.e. a thin EDL). Figure \ref%
{fig:char} shows $\varrho$ as a function of $r$. The calculations are made
from Eq.~\eqref{char0} using several values of \textrm{Da}. It can be seen
that both the size of the outer charged region and its integral charge grow
with Da. Recall that according to the definition (see \eqref{Da_def}) Da is
linear in $R$, which implies that the total charge of the outer cloud
depends on the particle size.

\begin{figure}[th]
\centering
\includegraphics[width=1\columnwidth ]{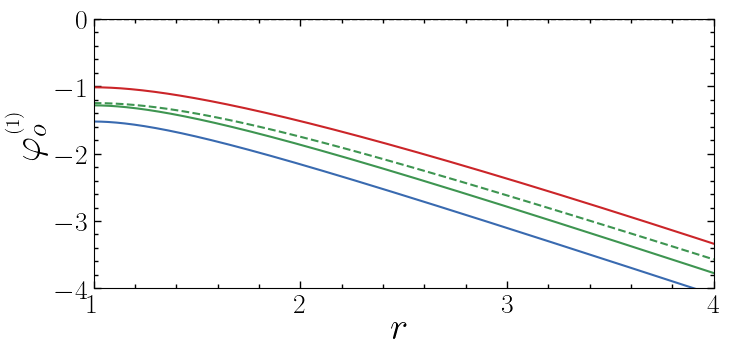} 
\caption{Outer disturbance potential $\varphi _{o}^{(1)}$ calculated using $\text{Da}=5$, $1$, $0$
(solid curves from top to bottom). Dashed curve shows calculations from Eq.~\eqref{smda} using $\mathrm{Da}=1$. }
\label{fig:fi1}
\end{figure}

In the general case solution to Eq.(\ref{du}) at a given Da that satisfies (%
\ref{bcs0}) and (\ref{bci0}) can be found only numerically, and note that it
does not depend on $\beta $. We solve it using standard Runge-Kutta
procedure with $\Delta r=0.01$. Figure \ref{fig:fi1} shows the radial
distribution of $\varphi _{o}^{(1)}$ computed for several $\mathrm{Da}$. It
can be seen that the $\varphi _{o}^{(1)}$-curves for all \textrm{Da} are
similar. The disturbance potential is negative for all $r$, and its
magnitude reduces on increasing \textrm{Da}.

At small $\mathrm{Da}$, the asymptotic solution to Eq.~(\ref{du}) can easily
be found by employing regular perturbation methods:%
\begin{equation}
\varphi _{o}^{(1)}=-r-\frac{1}{2r^{2}}+\frac{\mathrm{Da}}{2}\left( 1+\frac{1%
}{r^{3}}-\frac{3}{2r^{2}}\right) +O\left( \mathrm{Da}^{2}\right).
\label{smda}
\end{equation}%
Here the first two terms correspond to an inert particle (classical
electrophoresis). The third term is associated with the effect of (weak)
surface ion fluxes. The calculations from Eq.~\eqref{smda} made using $%
\mathrm{Da}=1$ are also included in Fig.~\ref{fig:fi1}. There is some
discrepancy from the numerical curve obtained using the same Da, but given
it is not really small, we have not expected \eqref{smda} to be accurate.
However, one can conclude that this equation is applicable as a first-order
estimate even when Da is finite.

\subsection{Inner solution}

The inner limits of the outer solution are the potential and concentration
at the particle surface. The surface potential $\varphi _{os}$ can be
obtained by substituting $r=1$ into Eqs.~\eqref{expa} and \eqref{eq:phi0_out}
that yields
\begin{equation}
\varphi _{os}=\varphi _{os}^{(0)}+\varepsilon \varphi _{os}^{(1)}\cos \theta,  \label{phis}
\end{equation}%
where
\begin{equation}
\varphi _{os}^{(0)}=\varphi _{o}^{(0)}\left( 1\right) =-\beta \ln \left( 1+%
\mathrm{Da}\right)   \label{phis0}
\end{equation}%
and
\begin{equation}
\varphi _{os}^{(1)}=\varphi _{o}^{(1)}\left( 1\right) .
\end{equation}%
Note that for $\mathrm{Da}\ll 1$ we can find $\varphi _{os}^{(1)}$ by
using Eq.(\ref{smda}):
\begin{equation}
\varphi _{os}^{(1)}=-\frac{3}{2}+\frac{\mathrm{Da}}{2}+O\left( \mathrm{Da}%
^{2}\right) .  \label{fsda}
\end{equation}

\begin{figure}[th]
\centering
\includegraphics[width=1\columnwidth ]{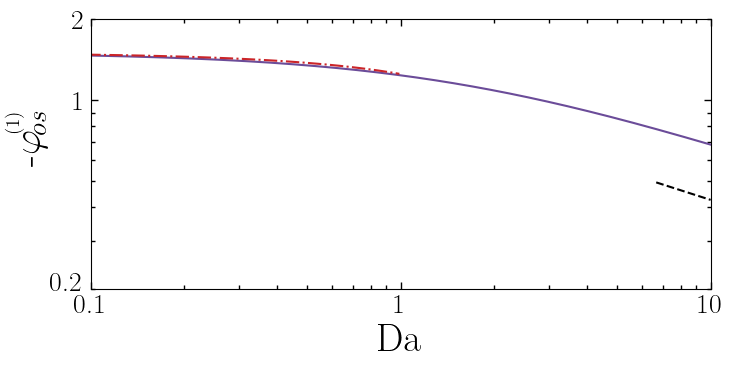} 
\caption{Disturbance potential $\varphi _{os}^{(1)}$ at the particle surface \emph{vs.} Da (solid curve).
Dashed-dotted curve shows calculations from asymptotic Eq.~\eqref{fsda}, dashed curve illustrates a $|\varphi _{os}^{(1)}|\propto\text{Da}^{-0.37}$ decay  at large values of $\text{Da}$.}
\label{fig:fi1s}
\end{figure}

Our calculations of $\varphi _{os}^{(1)}$ were based on differential equation \eqref{du}.
Figure \ref{fig:fi1s},  plotted in a log-log scale, shows numerical results for $-\varphi _{os}^{(1)}$  as a function of Da. It can be seen that on increasing Da the magnitude of $\varphi _{os}^{(1)}$ reduces. When $\Da \leq 1$ this decrease is excellently
fitted by Eq.~\eqref{fsda},  but for
large Damk\"{o}hler numbers the numerical fit gives $|\varphi _{os}^{(1)}|\propto\text{Da}^{-0.37}$. Below we will use this scaling
expression to calculate the contribution of the inner region to the particle velocity.

Correspondingly, substituting $r=1$ in Eq.~\eqref{c_sol} we obtain
\begin{equation}
c_{os}=c_{o}\left( 1\right) =1+\mathrm{Da}.  \label{c0}
\end{equation}

Since $\varphi _{os}^{(0)}$ and $c_{os}$ deviate from the bulk values, $%
c_{o}=1$ and $\varphi _{o}=0,$ it is natural to write the inner
dimensionless potential and concentrations in the form
\begin{equation}
\varphi =\varphi _{os}+\varphi _{i},\quad c_{i}^{\pm }=c_{os}\xi ^{\pm },
\end{equation}%
and to introduce a stretched coordinate $\rho =(r-1)/\lambda $, which
provides $\rho =O(1)$ within the inner region.

To satisfy the matching requirement at the zeroth order, at $\rho
\rightarrow \infty $ we impose
\begin{equation}
\xi ^{\pm }=1,\quad \varphi _{i}=0,  \label{ou_l}
\end{equation}%
which, taking into account \eqref{phis}-\eqref{c0}, indicates that $\varphi _{i}$ is
indeed the solution of the original problem. Note that conditions %
\eqref{ou_l} are identical to those for an inert particle.

The boundary condition for ion fluxes (\ref{bc_1}) at $\rho =0$, rewritten
in terms of $\rho ,$ reads
\begin{equation*}
\lambda ^{-1}c_{os}\left( \partial _{\rho }\xi ^{\pm }\pm \xi ^{\pm
}\partial _{\rho }\varphi _{i}\right) =-\mathrm{Da}\frac{1\mp \beta }{2}.
\end{equation*}%
Therefore, we obtain $\partial _{\rho }\xi ^{\pm }\left( 0\right) \pm \xi
^{\pm }\partial _{\rho }\varphi _{i}\left( 0\right) \sim \lambda \ll 1,$
which indicates that the fluxes in the inner region can safely be neglected.
We can then conclude that the ion concentration fields satisfy the Boltzmann
distribution,%
\begin{equation}
\xi ^{\pm }=\exp \left( \mp \varphi _{i}\right) ,  \label{Bol}
\end{equation}%
and that the potential obeys the Poisson-Boltzmann equation
\begin{equation}
\partial _{\rho \rho }\varphi _{i}=c_{os}\sinh \varphi _{i}.  \label{PB}
\end{equation}%
At the surface, $\rho =0$, the boundary condition to (\ref{PB}) reads
\begin{equation}
\varphi _{i}(0)=\phi _{s}-\varphi _{os}.  \label{phi_s}
\end{equation}%
Therefore, the inner-problem equations and boundary conditions are similar
to those for inert particle, but the bulk values, $c_{o}=1$ and $\varphi =0,$
should be replaced by $c_{os}$ and $\varphi _{os}.$

The nonlinear Poisson-Boltzmann equation (\ref{PB}) in spherical coordinates
can in general be solved only numerically. Its linearization is justified
provided $|\varphi _{i}^{(0)}|\leq 1$ and allows one to derive an explicit
analytical solution for spherically symmetric part of the field%
\begin{equation*}
\varphi _{i}^{(0)}=\left( \phi _{s}-\varphi _{os}^{(0)}\right) \exp \left(
-\rho c_{os}^{1/2}\right)
\end{equation*}%
From Eqs.~\eqref{eq:phi0_out}, \eqref{phis0} and \eqref{c0} it follows then
that the uniformly valid potential profile is given by
\begin{eqnarray}
\varphi ^{(0)} &=&\varphi _{o}^{(0)}\left( r\right) +\left( \phi
_{s}-\varphi _{os}^{(0)}\right) \exp \left( -\rho c_{os}^{1/2}\right)
\label{psih} \\
&=&-\beta \ln \left( 1+\frac{\mathrm{Da}}{r}\right)   \notag \\
&+&\left[ \phi _{s}+\beta \ln \left( 1+\mathrm{Da}\right) \right] \exp \left[
-\frac{(r-1)\left( 1+\mathrm{Da}\right) ^{-1/2}}{\lambda }\right]   \notag
\end{eqnarray}%
In the limit $\mathrm{Da}\rightarrow 0$ this equation becomes identical to %
\eqref{eq:DH}. Thus we recover the Debye-H\"{u}ckel formula.

\begin{figure}[th]
\centering
\includegraphics[width=1\columnwidth ]{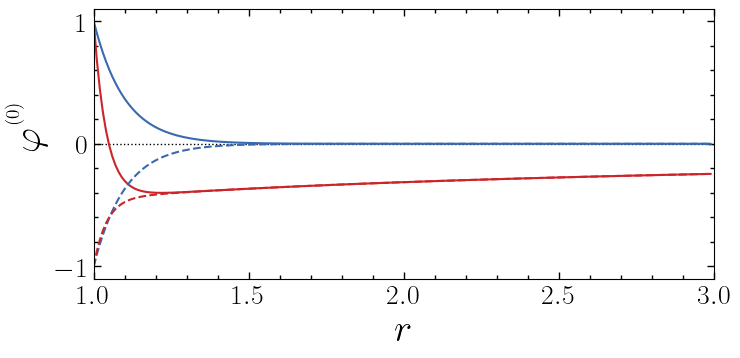} 
\caption{Potential $\protect\varphi^{(0)}$ as a function of distance $r$
calculated using $\protect\lambda=0.1$ and $\protect\beta=0.25$. Solid curves
from top to bottom show calculations for $\text{Da}=0$ and 5 made using $%
\protect\phi_s=1$. The same for $\protect\phi_s = -1$ is shown by dashed
curves. }
\label{fig:fi0}
\end{figure}

In Fig.~\ref{fig:fi0} we plot the radial distribution of $\varphi $ for
inert ($\mathrm{Da}=0$) and catalytic ($\mathrm{Da}=5$) particles.
Calculations are made from Eq.~\eqref{psih} using $\lambda =0.1$ and
(positive) $\beta =0.25$ implying that cations diffuse faster than anions. It
can be seen that near inert particles $|\varphi _{i}^{(0)}|$ decays rather
fast down to zero reflecting the emergence of the enriched by counter-ions
EDL. For catalytic particles the EDL becomes thinner, but is supplemented by
an outer diffuse layer that extends to large distances. In this layer the
negative (for positive $\beta $) radial potential tends to zero very slowly
and remains significant at $r=O(1)$. Note that the outer potential is
insensitive to $\phi _{s}$. This implies that the potential near the
positively charged particle reverses its sign at some $r$ within the EDL
from positive to negative. We emphasize, however, that this does not mean that the
part of the EDL becomes enriched by co-ions since in our case the Boltzmann
distribution applies only for $\varphi _{i}$, but not to $\varphi ^{(0)}$ in
whole.

\section{Particle velocity}\label{sec:velocity}

Now, we are in a position to answer our original question, which is to
obtain the expression for a migration speed of the catalytic particle.

The velocity of a freely moving in the $x-$direction particle can be
determined by using the reciprocal theorem~\cite%
{teubner1982,Masoud_Stone_2019}
\begin{equation}
u = -\frac{1}{6\pi }\int_{\mathcal{V}}\mathbf{f}\cdot \left( \mathbf{v}_{St}-%
\mathbf{e}_{x}\right) d\mathcal{V}.  \label{f_exp}
\end{equation}%
The integral is evaluated over the whole fluid volume $\mathcal{V}$ and $%
\mathbf{v}_{St}\left( \mathbf{r}\right) $ represents the velocity field for
the particle of the same radius that translates with the velocity $\mathbf{e}%
_{x}$ in a stagnant fluid (Stokes solution), so that%
\begin{equation}
\mathbf{v}_{St}-\mathbf{e}_{x}=u_{r}\cos \theta \mathbf{e}_{r}-u_{\theta
}\sin \theta \mathbf{e}_{\theta },  \label{v_st}
\end{equation}%
\begin{equation*}
u_{r}=\frac{3}{2r}-\frac{1}{2r^{3}}-1,\quad u_{\theta }=\frac{3}{4r}+\frac{1%
}{4r^{3}}-1.
\end{equation*}%
The contributions of the outer and the inner regions into the volume force
in \eqref{v_rec} can be decomposed
\begin{equation}
\mathbf{f}=-\Delta \varphi _{o}\mathbf{\nabla }\varphi _{o}-\Delta \varphi
_{i}\mathbf{\nabla }\left( \varphi _{o}+\varphi _{i}\right) -\Delta \varphi
_{o}\mathbf{\nabla }\varphi _{i}.  \label{fF}
\end{equation}%
The first and second terms are associated with the outer and inner regions,
correspondingly. We remark that $\Delta \varphi _{o}$ is small compared to $%
\Delta \varphi _{i}$ inside the EDL, while $\mathbf{\nabla }\varphi _{i}$ is
negligible in the outer domain. Therefore, the last term in \eqref{f_exp}
can safely be neglected in both regions and the particle velocity can be
presented as a superimposition of the contributions of the inner and the
outer regions~\cite{asmolov2022self},
\begin{equation}
u = \varepsilon \left( u_{i} + u_{o}\right).  \label{v_full}
\end{equation}%
The first term in Eq. (\ref{v_full}) is due to the slip velocity at the
outer edge of the inner region (see Appendix~\ref{app})%
\begin{equation}
\mathbf{v}_{\mathtt{s}}=\varepsilon \left( \phi _{s}-\varphi
_{os}^{(0)}\right) \nabla _{\mathtt{s}}\left( \varphi _{os}^{(1)}\cos \theta
\right).  \label{slip}
\end{equation}%
Here $\nabla _{\mathtt{s}}$ is the gradient operator along the particle
surface and the first multiplier in (\ref{slip}) is the potential drop in the
inner region. Recall that $\varphi _{os}^{(1)}$ for small Da is given by \eqref{fsda}.

The contribution of the inner region to the particle velocity is the average
slip velocity over the particle surface $\mathcal{S}$ \cite%
{prieve1984motion,anderson1989colloid}:
\begin{equation}
u_{i}=-\frac{1}{4\pi \varepsilon }\int_{\mathcal{S}}\left( \mathbf{v}_{%
\mathtt{s}}\cdot \mathbf{e}_{x}\right) d\mathcal{S}.  \label{ui0}
\end{equation}%
In Appendix~\ref{app} we perform the integration of \eqref{ui0}  and obtain
\begin{equation}
u_{i}=-\frac{2}{3}\left( \phi
_{s}-\varphi _{os}^{(0)}\right) \varphi _{os}^{(1)}.  \label{ui}
\end{equation}%
Thus, $u_{i}$ is determined by the product of the potential drop $\phi
_{s}-\varphi _{os}^{(0)}$ across the inner region  and the disturbance potential $\varphi _{os}^{(1)}$. The drop of potential in the EDL depends on Da, and
the potential $\varphi _{os}^{(0)}$ is proportional to $\beta $ (see
Eq. \eqref{phis0}) while $\varphi _{os}^{(1)}$ is independent of $\beta $.
Hence, the velocity $u_{i}$ is a linear function of $\beta$ that also depends nonlinearly on the Damk\"{o}hler number. It can in general be calculated only numerically, but using Eqs. (\ref{phis0}) and (\ref{fsda}) one can obtain the linear approximation
for $\mathrm{Da} \ll 1$:%
\begin{equation}
u_{i} \simeq \phi_{s} +  \mathrm{Da} \left(\beta - \dfrac{\phi_s}{3} \right).  \label{uism}
\end{equation}
This equation clarifies that only for $\mathrm{Da} = 0$ or $\beta = \phi_s/3$ the inner region contribution will be the same as for inert particles, \emph{i.e.} $u_i = \phi_s$. We also remark that it is not necessary for a catalytic particle to be charged to induce $u_i$. Indeed, $u_i = \beta \mathrm{Da}$ as $\phi_s = 0$.

\begin{figure}[th]
\centering
\includegraphics[width=1\columnwidth ]{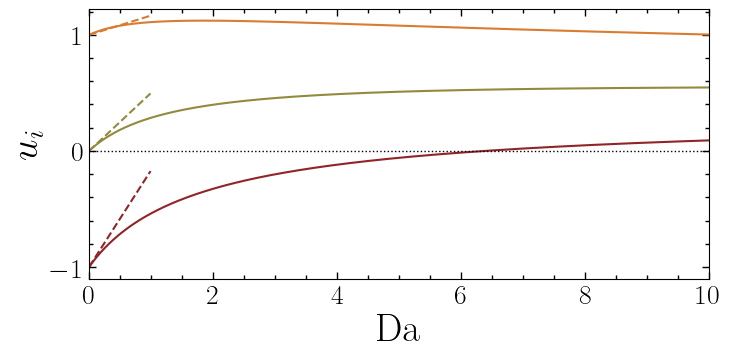} 
\caption{Contribution of the inner region to the particle velocity \emph{vs.} Da
calculated using $%
\protect\beta =0.5$ and $\protect\phi _{s}=1$, $0$ and $-1$ (solid curves from top to bottom). Dashed lines show the same, but are calculated from Eq.~\eqref{uism}.}
\label{fig:vi}
\end{figure}

To examine a contribution of the inner flows given by Eq.~\eqref{ui}, we fix $\beta = 0.5$ and plot
 $u_{i}$  \emph{vs.} Da for several $\phi_s$ in Fig.~\ref{fig:vi}. Also included are calculations from Eq.~\eqref{uism}.  It can be seen that for an inert particle  $u_{i} = \phi _{s}$, which is the consequence of $\varphi _{os}^{(0)}=0$ and $\varphi _{os}^{(1)}=-3/2$. However, at $\mathrm{Da} \neq 0$ the value of $u_{i}$ becomes different from $\phi_s$. It follows from Eq.~\eqref{eq:phi0_out} that $\varphi _{os}^{(0)}$ is
negative for positive $\beta$. This implies that in the case of positive $\phi_s$ the potential drop, $\phi
_{s}-\varphi _{os}^{(0)}$, in \eqref{ui} is positive and augments with Da. However, the magnitude of $\varphi _{os}^{(1)}$ reduces with Da.
As a result, the function $u_i(\mathrm{Da})$ takes its maximum value at $\Da \simeq 2$ (with our parameters).
 If $\phi_s$ is negative, $\phi _{s}-\varphi _{os}^{(0)}$ reverses its sign at $\Da \simeq 6$, so is $u_i$.
As a side note, at extremely large \textrm{Da} all the curves should collapse (very slowly) to a single one since
 $\varphi _{os}^{(0)} \varphi _{os}^{(1)}\propto \ln (\mathrm{Da})\mathrm{Da}^{-0.37}$.

\begin{figure}[th]
\centering
\includegraphics[width=\columnwidth ]{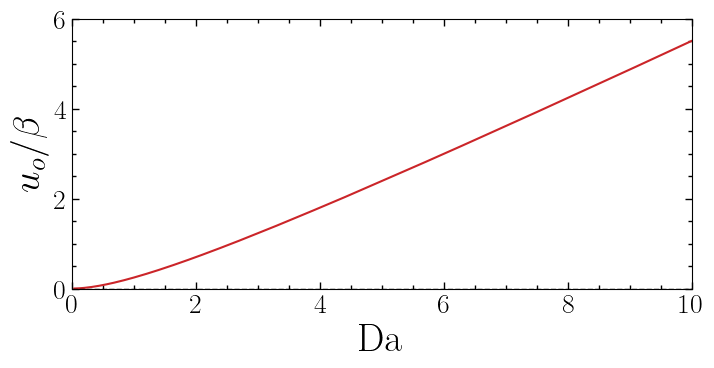} 
\caption{Contribution of the outer region to the particle velocity \emph{vs.} Da
(solid curve). }
\label{fig:vo}
\end{figure}

The second term in \eqref{v_full} that is associated with an outer region
can be written as \cite{asmolov2022self}
\begin{equation}
u_{o}=-\frac{1}{6\pi }\int_{\mathcal{V}}\mathbf{f}_{o}\cdot \left( \mathbf{v}%
_{St}-\mathbf{e}_{x}\right) d\mathcal{V},  \label{uo}
\end{equation}%
where the integral is taken over the outer region and
\begin{eqnarray*}
\mathbf{f}_{o} &\simeq &-\left[ \Delta \varphi _{o}^{(0)}\mathbf{\nabla }%
\left( \varphi _{o}^{(1)}\cos \theta \right) +\Delta \left( \varphi
_{o}^{(1)}\cos \theta \right) \mathbf{\nabla }\varphi _{o}^{(0)}\right]  \\
&=&-\beta \left( \frac{\partial _{r}c_{o}}{c_{o}}\right) ^{2}\left( 2\cos
\theta \partial _{r}\varphi _{o}^{(1)}\mathbf{e}_{r}-\frac{\sin \theta }{r}%
\varphi _{o}^{(1)}\mathbf{e}_{\theta }\right) .
\end{eqnarray*}%
By integrating Eq. (\ref{uo}) over $\theta $ one can derive%
\begin{equation}
u_{o}/\beta = I_{o},  \label{io}
\end{equation}%
where
\begin{equation}
I_{o}=\frac{4}{9}\int_{1}^{\infty }r^{2}\left( \frac{\partial _{r}c_{o}}{%
c_{o}}\right) ^{2}\left( \partial _{r}\varphi _{o}^{(1)}u_{r}+\frac{\varphi
_{o}^{(1)}u_{\theta }}{r}\right) dr  \label{io1}
\end{equation}
depends on the Damk\"{o}hler number. To integrate Eq.~(\ref{io1}) numerically, we define a new variable $%
s=r^{-1}$. The integral is then computed on uniform grid in $s$ using $%
N_{s}=200$. The results of these computations are included in  Fig.~\ref{fig:vo}, where
$u_{o}/\beta $ is plotted as a function of Da. We see that for all
$\mathrm{Da} \neq 0$ integral (\ref{io1}) is positive, indicating that $u_{o}$ and $\beta $ are of the same sign. It demonstrates
a quadratic growth for small values of $\mathrm{Da}$ since $\partial
_{r}c_{o}\propto \mathrm{Da}$. For larger Damk\"{o}hler numbers, it augments
linearly and exceeds in several times the contribution of the inner region $%
u_{i}$, since the size of the charged outer region grows with $\mathrm{Da}$ (see Fig.~\ref{fig:char}).

We can now find the migration speed of the particle and zeta potential, which are determined both by the surface potential $\phi _{s}$ and by the charged outer region.
By using Eqs.~\eqref{ui} and \eqref{io} we obtain%
\begin{equation}
u=\varepsilon \left[ u_{\phi } +\beta u_{\beta
} \right],
\label{utot}
\end{equation}
where
\begin{equation}
u_{\phi }=-\frac{2}{3}\phi _{s}\varphi _{os}^{(1)},\quad u_{\beta }=-\frac{2%
}{3}\ln \left( 1+\mathrm{Da}\right) \varphi _{os}^{(1)}+I_{o},
\label{uphi}
\end{equation}
and note that both $u_{\phi }$ and $u_{\beta }$ depend on $\mathrm{Da}$. Correspondingly, the dimensionless zeta potential of the particle, $\zeta = \mathrm{e} Z/(k_B T)$, is given by
\begin{equation}\label{eq:zeta}
 \zeta = \dfrac{u}{\varepsilon},
\end{equation}
and recall that the measured $\zeta$ is traditionally used  to infer the surface potential of inert particles. Below we shall see that for
catalytic particles $\zeta \neq \phi_s$, even if $\beta =0$ and, consequently,  the second term in \eqref{utot} vanishes. This becomes apparent, if we remember that $\varphi _{os}^{(1)} \neq -3/2$ for $\mathrm{Da} \neq 0$  (see Fig. \ref{fig:fi1s}).

\begin{figure}[th]
\centering
\includegraphics[width=1\columnwidth ]{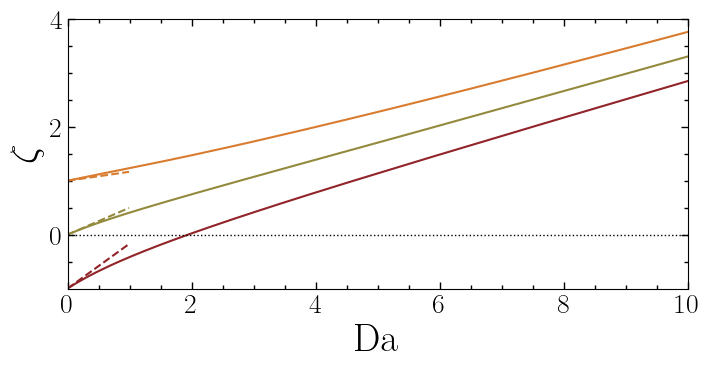} 
\caption{Zeta potential as a function of Da calculated using $\protect\phi %
_{s}=1 $, $0$, $-1$ (solid curves from top to bottom) and $\protect\beta =0.5$. Dashed lines show $\zeta = u_i$ calculated from Eq.~\eqref{uism}.}
\label{fig:vt}
\end{figure}

By varying Da at fixed $\phi_s$ it is possible to obtain the $\zeta$-curves shown in Fig.~\ref{fig:vt}. The calculations are made using $\beta = 0.5
$ and $\phi_s = 1, 0,$ and $-1$. It is well seen that $\zeta = \phi_s$ only for inert particles of $\mathrm{Da}= 0$. On increasing Da all curves
grow monotonically. For small Damk\"{o}hler numbers $\zeta$ is reasonably well approximated by Eq.~\eqref{uism}, confirming that electrophoresis is controlled mostly by the inner region. However, this asymptotic expression is inapplicable, if Da becomes finite, since there emerges a significant contribution of the outer cloud. If the particle is positively charged, its $\zeta$ remains positive (indicating a migration along the field $E$) and
increases with Da. For instance, $\zeta \simeq 3$ when $\mathrm{Da}\simeq 8$, indicating that the catalytic particle migrates \emph{ca.} 3 times faster than
inert of the same (positive) surface potential. The neutral particle would be immobile when inert since the EDL is not formed. However, the
catalytic neutral particle migrates in the direction of the external field like it is positively charged. At large Da its zeta potential becomes
rather high, although remains smaller than of the catalytic particle with positive $\phi_s$. If $\phi_s$ is negative, on increasing Damk\"{o}hler
number the negative $\zeta$ reduces and vanishes at some Da. On increasing Da further $\zeta$ becomes positive and increases. In other words, the
particle first migrates opposite to the field and retards its speed, but then reverses the direction of motion and accelerates with Da.

\begin{figure}[th]
\centering
\includegraphics[width=1\columnwidth ]{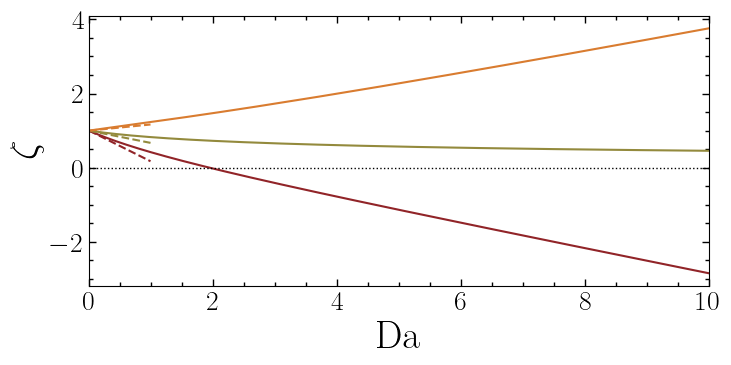} 
\caption{Zeta potential as a function of Da calculated using $\protect\beta %
= 0.5 $, $0$, $-0.5$ (from top to bottom) and $\protect\phi_s =1$. Dashed lines show $\zeta = u_i$ calculated from Eq.~\eqref{uism}.}
\label{fig:vtbeta}
\end{figure}

Another parameter that controls $\zeta$ is $\beta$ given by Eq.~\eqref{bet}. We recall that the results shown in Fig.~\ref{fig:vt} refer to positive $\beta$, which implies that $D^+$ exceeds $D^-$. If $\beta$ is negative, the zeta potential can be found by using the identity $
\zeta \left( \phi _{s},\beta \right) =-\zeta \left(-\phi _{s},-\beta \right) $ that follows from Eqs.~(\ref{utot}) and (\ref{uphi}). This implies
that for say $\phi_s = 1$, but $\beta = -0.5$, the $\zeta-$curve will simply represents a mirror image (relative to $\zeta = 0$) of the
curve in Fig.~\ref{fig:vt} obtained for $\phi_s = -1$  and $\beta = 0.5$. This is illustrated in Fig.~\ref{fig:vtbeta}, where we also included the curve for $\beta = 0$. In this specific case $\zeta$ reduces with Da and tends to zero at $\mathrm{Da} \to \infty$, indicating that the positively charged colloid particle does not migrate under an applied field. This follows by rewriting Eq.~\eqref{utot} as
$\zeta =-2 \phi_s \varphi _{os}^{(1)}/3$ (since its second term vanishes) and recognizing that $\varphi _{os}^{(1)}$ is negative for any Da  (see Fig. \ref{fig:fi1s}), we conclude that $\zeta$ and $\phi_s$ are always of the same sign. However, since $\varphi _{os}^{(1)}$ reduces with \Da, the zeta potential is lowered, although very slowly.

\section{Concluding remarks}\label{sec:conclusion}

We developed a theory of electrophoresis of catalytic particles
that uniformly release ions. Attention was focused on the case of a weak field and large compared to a diffuse layer particles.  In other words, we have addressed a  situation that is traditionally described by the famous Smoluchowski theory~\cite{smoluchowski.m:1921}. The main conclusion of this theory is that the  velocity of electrophoretic migration of colloid particle is determined by the zeta potential $\zeta$ of its surface, which in turn is equal to the surface potential $\phi_s$, if the no-slip boundary condition is postulated. This, however, refers to inert particles only. Here we have shown that the Smoluchowski picture can be dramatically changed if particles become catalytic. Our model, which is probably the simplest realistic model for an ion release from the particle surface that one might contemplate, provides considerable insight into electrophoretic migration of large catalytic particles and their zeta potential. The latter, beside the surface potential, now depends on the surface flux (quantified by the Damk\"{o}hler number Da) and on the  parameter $\beta$, which is a characteristic of the difference in ion diffusivity.

We derive that when $\mathrm{Da} = 0$ the electrophoretic velocity of colloid particles is determined solely by their surface potential, \emph{i.e.} that the zeta and surface potentials are equal. Thus, we recover the classical Smoluchowski result. However, $\zeta \neq \phi_s$ when $\mathrm{Da} \neq 0$.
For small values of Da we show that $\zeta$, and, consequently, the direction of migration is controlled mostly be $\phi_s$, although affected both by Da and $\beta$. For large Da the magnitude of the zeta potential is only slightly affected  by $\phi_s$ and is mostly set by the value of Da. In turn, the direction of electrophoretic migration, \emph{i.e.} the sign of $\zeta$,  becomes insensitive to the sign of the surface potential, and is controlled by the sign of $\beta$. If $\beta$ is positive, which implies that the diffusion coefficient of cations exceeds that of anions, all particles  migrate along the applied field by creating an (erroneous)  impression that they are positively charged. By contrast, negative $\beta$ always induces the migration against the field like it would be for negatively charged inert particles. We stress that even neutral particles (of $\phi_s = 0$) can be of quite a large zeta potential, provided Da is large and $\beta \neq 0$. If $\beta = 0$, the zeta potential will be always of the same sign as $\phi_s$, and its magnitude monotonically decreases down to zero with Da. In other words, in this particular case of equal ion diffusivities, their release can only slow down the particle or even stop it. Note that this result is different from derived for the diffusiophoretic migration, which for catalytic particles will be the same as of inert ones, if $\beta$ vanishes~\cite{asmolov.es:2024}.

These aspects of our work warrant further comments. For the case of inert particles we have emphasized that $\zeta$ is completely independent on the dimension of the particle, when large enough. This fact is, of course, well known and clarified yet in early literature~\cite{overbeek}. The dependence of zeta potential of large catalytic particles on $\mathrm{Da}\propto R$ (see Eq.~\eqref{Da_def}) we have found here implies that their $\zeta$ is radius-dependent even in the thin EDL limit. This open an avenue for novel  concepts of particle separation by size based on a linear electrophoresis. We have also stressed the connection between the direction of electrophoretic migration of catalytic particles and the parameter $\beta$, which is another feature compared to inert particles. Recall that electrophoresis of the latter does not depend on the diffusivity of electrolyte ions. At first sight it is somewhat surprising that (at sufficiently large the Damk\"{o}hler numbers) the direction of electrophoretic migration is set by the sign of $\beta$, but not of the surface potential. However, it becomes self-evident from the treatment presented here that catalytic particles should, beside the standard EDL, form a large outer cloud due to diffusion of released ions. However, it becomes self-evident from the treatment presented here that catalytic particles should, beside the standard EDL, form a large outer cloud since released ions  diffuse outside the EDL. Although weakly charged, this cloud extends to distances of the order of the particle size. At large Da its contribution to the particle velocity becomes dominant, so the direction of electrophoretic migration will be determined solely by the sign of the outer cloud charge or, equivalently, by the sign of $\beta$.

The existence of the outer cloud around catalytic particles implies that ionic concentrations exceed the bulk value at larger distances from their surface compared to the case of inert colloids, and that the electroneutrality breaks at much larger distances than the EDL dimension. Our analysis has clarified the repercussions of these for a single-particle electrophoresis. For dilute suspensions of inert particles such a single-particle picture is usually enough to treat the problem since the flow is confined inside a thin EDL only. If we deal with suspensions of catalytic particles, the flow might be generated at much larger distances from the surface leading to long-range interactions of particles, and, hence, to their collective electrophoretic motion. It appears that the phenomenon of extremely long-range interaction of releasing ions porous particles~\cite{feldmann.d:2020} is consistent with  our theoretical description of an outer cloud, but we are unaware of any work that addressed the issue of a collective  electrophoretic motion in such systems. It would be of considerable interest to study this both theoretically and experimentally.

Summarizing, the extension of the theory to the case of catalytic particles has provided a new insight into the origin and physics of electrophoretic propulsion, and into the various factors that determine the zeta potential of colloid particles.

\appendix

\section{Derivation of expression for $u_i$ using the reciprocal theorem}\label{app}

We evaluate the contribution of the inner region to the particle velocity,
i.e. the contribution of the second term in (\ref{fF}) to integral (\ref%
{f_exp}). The auxiliary velocity field (\ref{v_st}) can be presented in
terms of stretched variable $\rho $ as%
\begin{equation*}
\mathbf{v}_{St}-\mathbf{e}_{x}\simeq -\frac{3}{2}\rho ^{2}\lambda ^{2}\cos
\theta \mathbf{e}_{r}+\frac{3}{2}\rho \lambda \sin \theta \mathbf{e}_{\theta
},
\end{equation*}%
i.e. the velocity is small in this region. However, the $\rho $ derivatives
of $\varphi _{i}$ are large, so that, $\Delta \varphi _{i}\simeq \lambda
^{-2}\partial _{\rho \rho }\varphi _{i},\ \mathbf{\nabla }\varphi
_{i}=\lambda ^{-1}\partial _{\rho }\varphi _{i}\mathbf{e}_{r}+\partial
_{\theta }\varphi _{i}\mathbf{e}_{\theta }.$ As a result the integral over
the EDL is finite:%
\begin{widetext}
\begin{equation}
u_{i}=-\frac{1}{6\pi }\int_{\mathcal{V}}\mathbf{f}\cdot \left( \mathbf{v}%
_{St}-\mathbf{e}_{x}\right) d\mathcal{V}=-\frac{1}{2}\int\limits_{0}^{\pi
}\int\limits_{0}^{\infty }\partial _{\rho \rho }\varphi _{i}\left[ \partial
_{\rho }\varphi _{i}\rho ^{2}\cos \theta -\partial _{\theta }\left( \varphi
_{os}+\varphi _{i}\right) \rho \sin \theta \right] d\rho \sin \theta d\theta
.  \label{vfi2}
\end{equation}

Integrating the first term in the square brackets over $\rho $ by parts, we obtain:%
\begin{equation}
-\frac{\sin 2\theta }{2}\int\limits_{0}^{\infty }\partial _{\rho \rho
}\varphi _{i}\partial _{\rho }\varphi _{i}\rho ^{2}d\rho =-\frac{\sin
2\theta }{2}\int\limits_{0}^{\infty }\rho ^{2}d\left( \frac{\partial _{\rho
}\varphi _{i}}{2}\right) ^{2}=\sin 2\theta c_{os}\int\limits_{0}^{\infty
}\left( \cosh \varphi _{i}-1\right) \rho d\rho .  \label{f1ir}
\end{equation}%
The integration over $\rho $ of the  proportional to $\partial _{\theta
}\varphi _{os}$ term in (\ref{vfi2}) gives the electrophoretic slip velocity
\begin{equation}
\partial _{\theta }\varphi _{os}\int\limits_{0}^{\infty }\rho \partial
_{\rho \rho }\varphi _{i}d\rho =\partial _{\theta }\varphi
_{os}\int\limits_{0}^{\infty }\rho d\partial _{\rho }\varphi _{i}=-\partial
_{\theta }\varphi _{os}\left( \phi _{s}-\varphi _{os}^{(0)}\right) =\varphi
_{os}^{(1)}\sin \theta \left( \phi _{s}-\varphi _{os}^{(0)}\right) .
\label{f2r}
\end{equation}%

Using Poisson-Boltzmann equation (\ref{PB}) and integrating over $\theta$ by parts, we find for the proportional to $\partial _{\theta }\varphi _{i}$ term   in (\ref{vfi2}):
\begin{equation}
\rho c_{os}\int\limits_{0}^{\pi }\sinh \varphi _{i}\partial _{\theta
}\varphi _{i}\sin ^{2}\theta d\theta =\rho c_{os}\int\limits_{0}^{\pi }\sinh
\varphi _{i}\sin ^{2}\theta d\varphi _{i}=-\rho c_{os}\int\limits_{0}^{\pi
}\left( \cosh \varphi _{i}-1\right) \sin 2\theta d\theta .  \label{f2t2}
\end{equation}%
By taking integrals (\ref{f2t2}) over $\rho $
and (\ref{f1ir})
over $\theta$ we conclude that these two terms cancel. Thus the velocity $u_i$ is determined from (\ref{f2r}). Taking integral over
$\theta$ we derive the expression for catalytic particles
\begin{equation*}
u_{i}=-\frac{1}{2}\left( \phi _{s}-\varphi _{os}^{(0)}\right) \varphi
_{os}^{(1)}\int\limits_{0}^{\pi }\sin ^{3}\theta d\theta =-\frac{2}{3}\left(
\phi _{s}-\varphi _{os}^{(0)}\right) \varphi _{os}^{(1)}.
\end{equation*}%
Since for inert particles $\varphi _{os}^{(0)}=0$ and $\varphi _{os}^{(1)}=-3/2$, the latter equation reduces to $u_{i}=\phi _{s}.$ %
\end{widetext}

\begin{acknowledgments}

This work was supported by the Ministry of Science and Higher Education of the Russian Federation.
\end{acknowledgments}

\section*{DATA AVAILABILITY}

The data that support the findings of this study are available within the
article.

\section*{AUTHOR DECLARATIONS}

The authors have no conflicts to disclose.

\bibliographystyle{unsrt}
\bibliography{eph3}

\end{document}